\documentclass[11pt]{article}
\usepackage{epsfig,amsfonts,latexsym}
\usepackage{fullpage}
\newlength{\dblWidth}
\newlength{\sglWidth}
\newlength{\fpColGap}
\newlength{\fpColWidth}

\newcommand{\figurepair}[4]{
        \begin{figure}[thb]\mbox{%
        \makebox[\fpColWidth]{%
                #1
        }%
        \makebox[\fpColGap]{}%
        \makebox[\fpColWidth]{%
                #3
        }%
        }
        \mbox{%
        \parbox{\fpColWidth}{%
                #2
        }%
        \makebox[\fpColGap]{}%
        \parbox{\fpColWidth}{%
                #4
        }%
        }
        \end{figure}
}

\begin{document}
\bibliographystyle{apalike}
\vspace{2.1in}

\title{\bf Statistical Methods for Detecting Stellar Occultations by
Kuiper Belt Objects:  \\
the Taiwanese-American Occultation Survey}

\vspace{0.2in}
\author{
Chyng-Lan Liang\footnote{Department of Statistics, University of California, Berkeley. USA} \ %
 John A. Rice\footnote{Department of Statistics, University of California, Berkeley. USA} \ %
 Imke de Pater\footnote{Department of Astronomy, University of California, Berkeley. USA} \ %
 Charles Alcock\footnote{Department of Physics and Astronomy, University of Pennsylvania. USA} \\ %
 Tim Axelrod\footnote{Steward Observatory, University of Arizona. USA} \ %
 Andrew Wang\footnote{Academia Sinica, Institute of Astronomy and Astrophysics. Taiwan} \ %
  Stuart Marshall\footnote{Institute of Geophysics and Planetary
  Physics, Lawrence Livermore National Laboratory. USA}
  }
\maketitle
\begin{abstract}
The Taiwanese-American  Occultation Survey (TAOS) will detect objects
in the Kuiper Belt, by measuring the rate of occultations of stars by
these objects, using an array of three to four 50cm wide-field robotic
telescopes.  Thousands of stars will be monitored, resulting in
hundreds of millions of photometric measurements per night. To optimize
the success of TAOS, we have investigated various methods of gathering
and processing the data and developed statistical methods for detecting
occultations.  In this paper we discuss these methods. The resulting
estimated detection efficiencies will be used to guide the choice of
various operational parameters determining the mode of actual
observation when the telescopes come on line and begin routine
observations. In particular we show how real-time detection algorithms
may be constructed, taking advantage of having multiple telescopes.  We
also discuss a retrospective method for estimating the rate at which
occultations occur.
\end{abstract}

\section{Introduction}
Since the middle of the last century, there has been increasing
 speculation
 that a residual protoplanetary disk existed beyond
 Neptune consisting of a vast number of remnants of the accretional phase of the early evolution of the solar
system.  This belt is the source of most short--period  comets, those
with periods of 200 years or less \cite{edge}, \cite{kuiper},
\cite{fern}. Observational success was first achieved with the
discovery of 1992QB1
  \cite{jewitt5}. Major observational efforts since then have identified
about 500 objects, the largest having a diameter of about 900km.
Studies of the Kuiper belt have been reviewed in  \cite{weiss},
\cite{stern} and \cite{jewitt3}.

At 50AU (one AU is the average distance from the Sun to Earth, $1.49
\times 10^8$ km,), we can currently only directly observe objects
larger than around 100km, since smaller objects do not reflect
sufficient light. Thus, other methods are needed to detect smaller
objects, which are far greater in abundance. The idea of the
occultation technique \cite{bailey}, \cite{axel2} is simply the
following:
 One monitors the light from a sample of stars that
have angular sizes smaller than the expected angular sizes of Kuiper Belt
Objects (KBO's) we
hope  to detect. An
occultation is manifested by detecting the partial or total reduction in
the flux from one of the stars for a brief interval, when a KBO passes
between it and the observer. This technique will allow the detection of
objects with a radius of only a few kilometers and/or larger objects lying beyond 100AU, objects
which have thus far been undetectable by direct observation.

The Taiwanese-American Occultation Survey (TAOS), a
collaboration involving the
 Lawrence Livermore National
Laboratory (USA), Academia Sinica and National Central University (both
of Taiwan), will use this stellar occultation technique with an array
of
 three or four wide-field robotic telescopes to estimate the number of
KBOs of size greater than a few kilometers.
Each of these 50cm
telescopes will be pointed at the same 3 square degrees of the sky and
will record light from the same approximately 2,000 stars.
The telescope array
will be located in the Yu Shan (Jade Mountain) area of central Taiwan (longitude
$ 120 ^{\circ}$  50' 28'' E; latitude 23$^{\circ}$ 30' N).

 The
detection scheme will need to operate in real-time, as the data is
being gathered, in order to alert more powerful telescopes for
followup. It is anticipated that a large amount of data will be
generated on a nightly basis, yielding about 10,000 gigabytes of data
and $10^{10}-10^{12}$ occultation tests per year. The challenge is to
detect among these only small number of occultations, perhaps tens or
hundreds (the uncertainty in this number reflects our ignorance).

\section{TAOS Observing Scheme}

The imaging system at each telescope will consist of a
thermo-electrically cooled charge coupled device (CCD) camera with a
$2048 \times 2048$ pixel CCD detector
(pixel size is 2.89 arcseconds).

At the time the research presented here was conducted, the final
selection of star fields had not yet been made, other than that they
will be near the ecliptic plane (the plane of the Earth's orbit).
Although naively one might think that the denser star fields might be
more profitable for revealing occultations, the detection scheme has to
deal with crowding of stars and sky background. We will thus consider
both crowded and sparse fields (Figure \ref{twofield}).

\figurepair{\psfig{figure=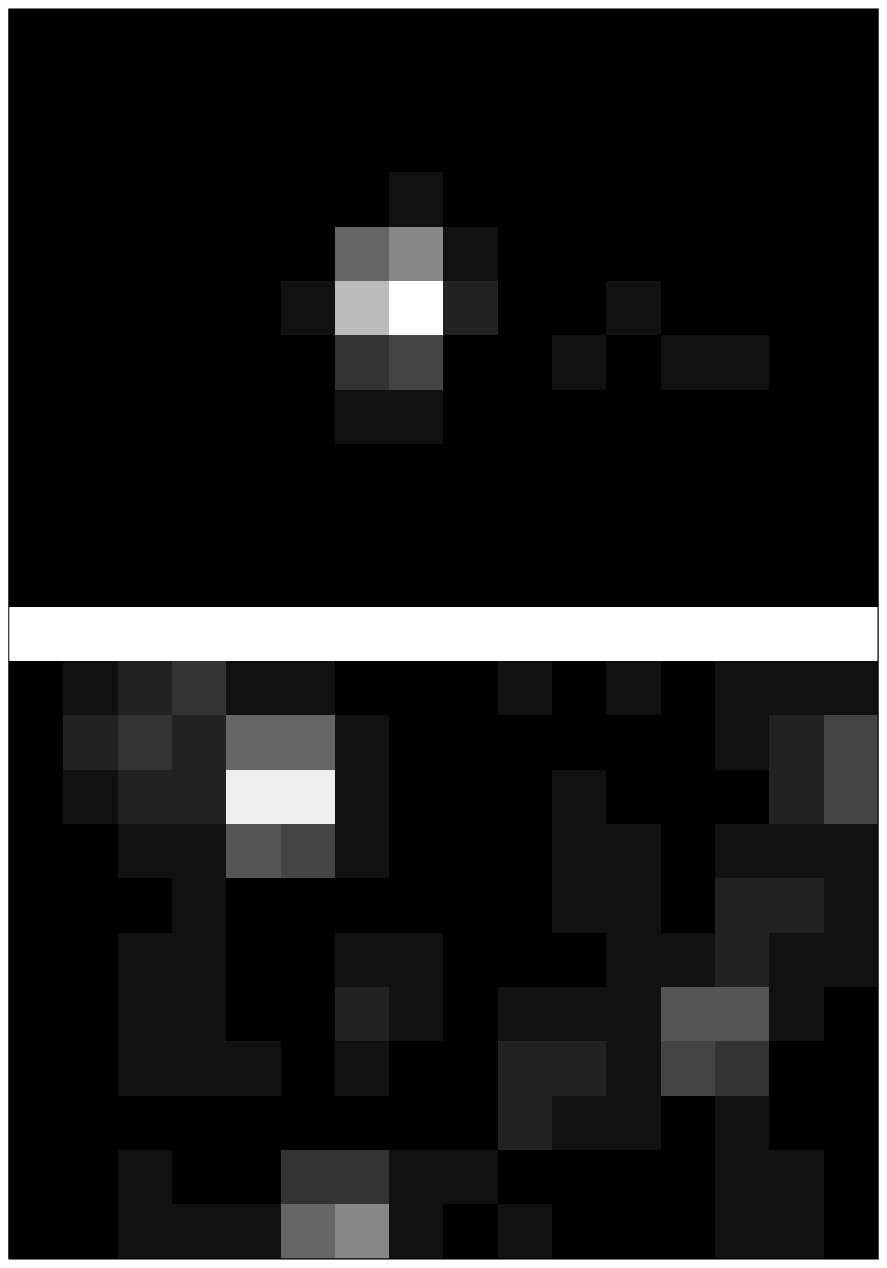,width=3.9in,angle=0}}
{\caption{Two 46 $\times$ 32 arcsec. sections of simulated images for
star fields (RA 4.905', Dec 29.275'') [Top] and (RA 9.61833', Dec
0.7250'') [Bottom]. } \label{twofield}}
{\psfig{figure=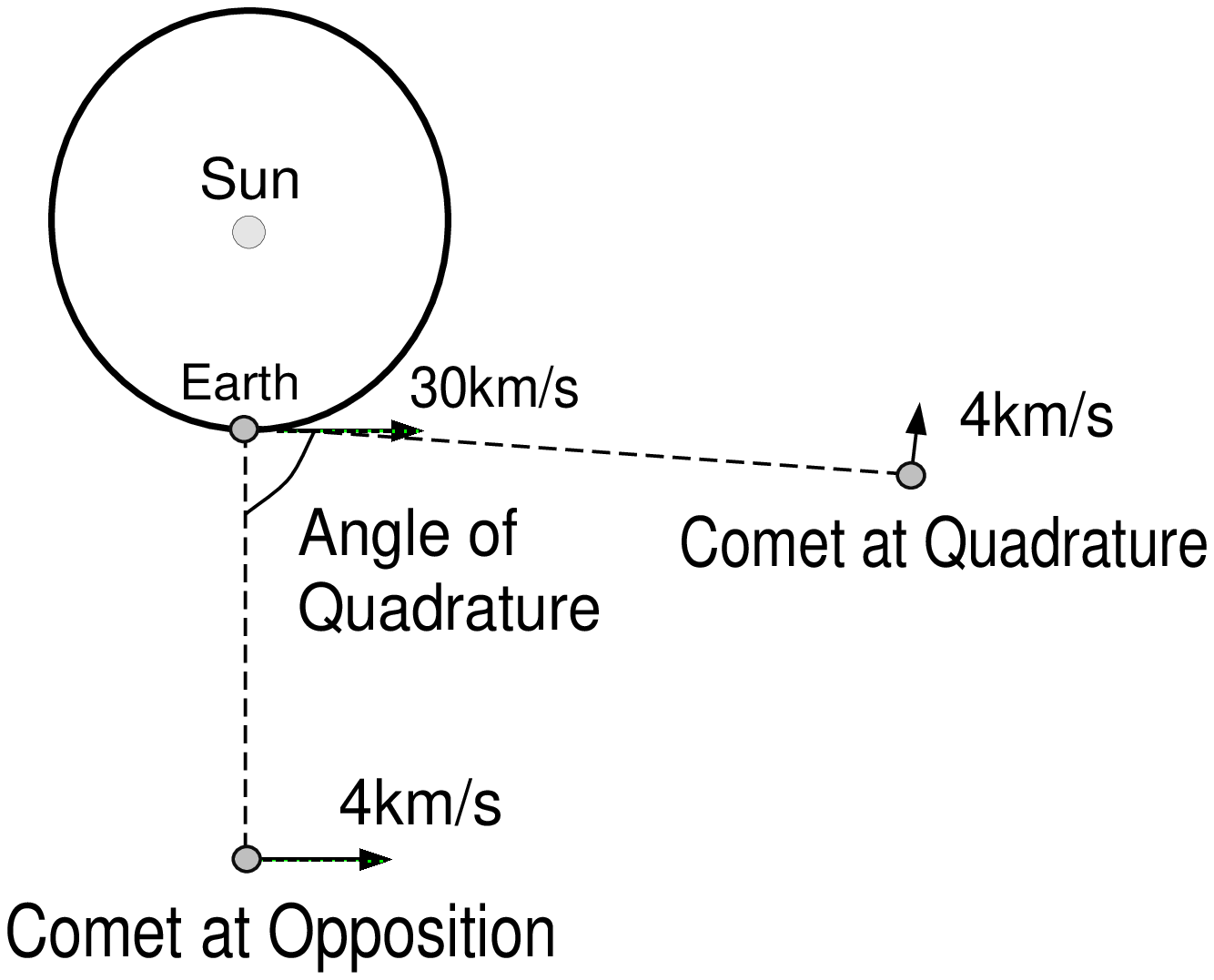,width=3.2in,angle=0}} {\caption{KBO at
opposition and at quadrature.} \label{quad}}

The angle of observation will affect the properties of the occultation,
since the relative velocity of the KBO with respect to Earth will
depend upon it. The KBO is said to be observed at {\em quadrature} when
the relative velocity is zero. The angle of quadrature depends upon the
KBO's heliocentric distance (the distance to the center of the Sun).
For a KBO at 50AU, the angle of observation at quadrature  is around 80
degrees. The KBO is said to be observed at {\em opposition} when the
relative velocity is at its maximum---for a KBO moving at 4km
$\rm{s}^{-1}$ (which is typical for an object at 50AU), this maximal
relative velocity is 26km $\rm{s}^{-1}$,  since Earth moves at about
30km s$^{-1}$. Figure \ref{quad} shows the KBO at opposition and at
quadrature. The relative velocity of an object in a circular orbit at
$r$ AU, at angle of observation $\varphi$, is given by:

\begin{equation}
RV(r,\varphi)=v_e \cos \varphi - v_e \sqrt{\left( \frac{r_e}{r} \right) \left(
1- \left( \frac{r_e}{r} \right)^2 \sin^2 \varphi \right)}
\end{equation}

\noindent where $v_e$ and $r_e$ are the velocity and heliocentric distance of the
Earth respectively. We will consider two relative velocities:
 20km s$^{-1}$ (near opposition) and 6km s$^{-1}$ (near quadrature).

\begin{figure}[htbp]
\centerline{\psfig{figure=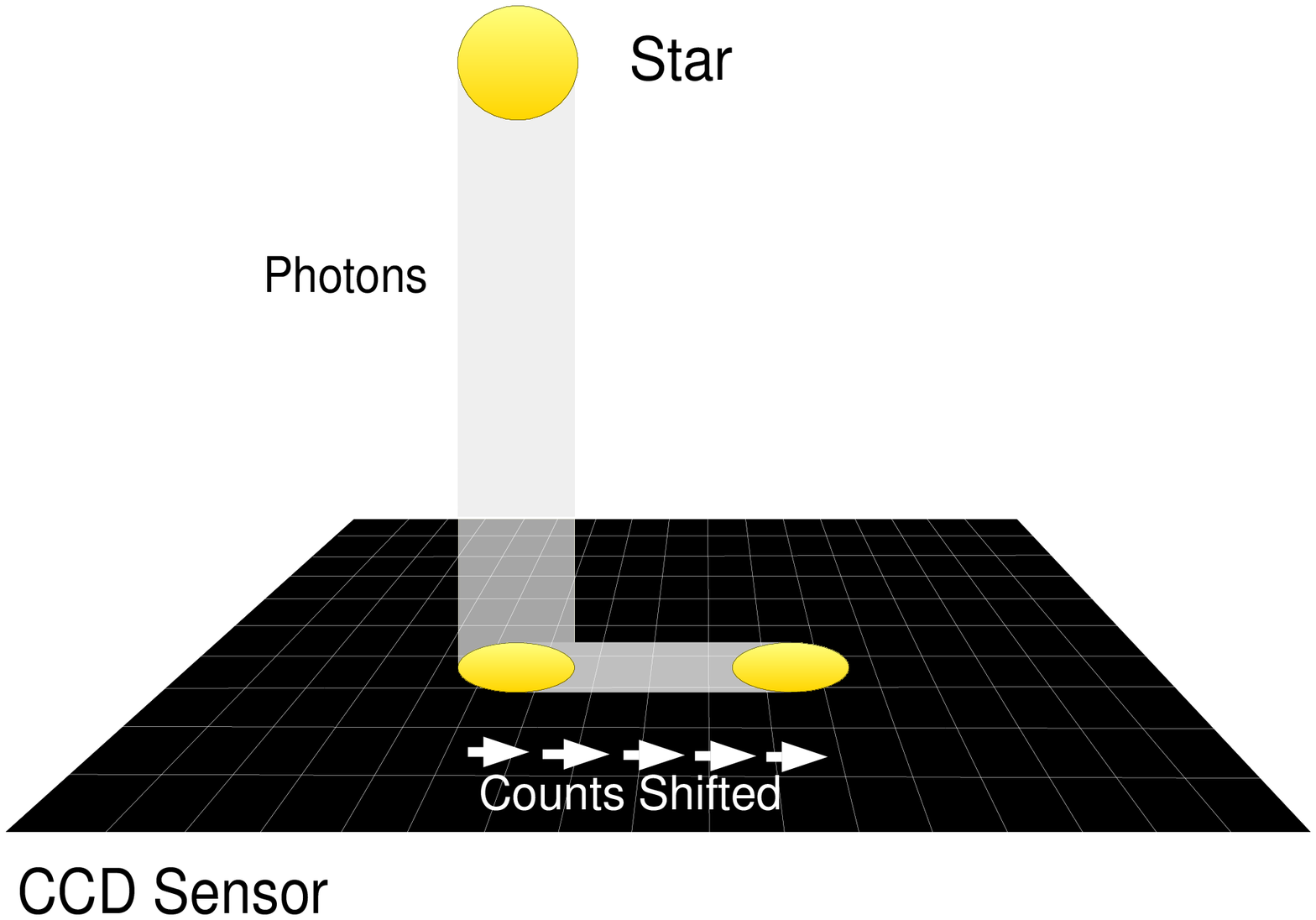,width=3.6in,angle=0}}
\centerline{\psfig{figure=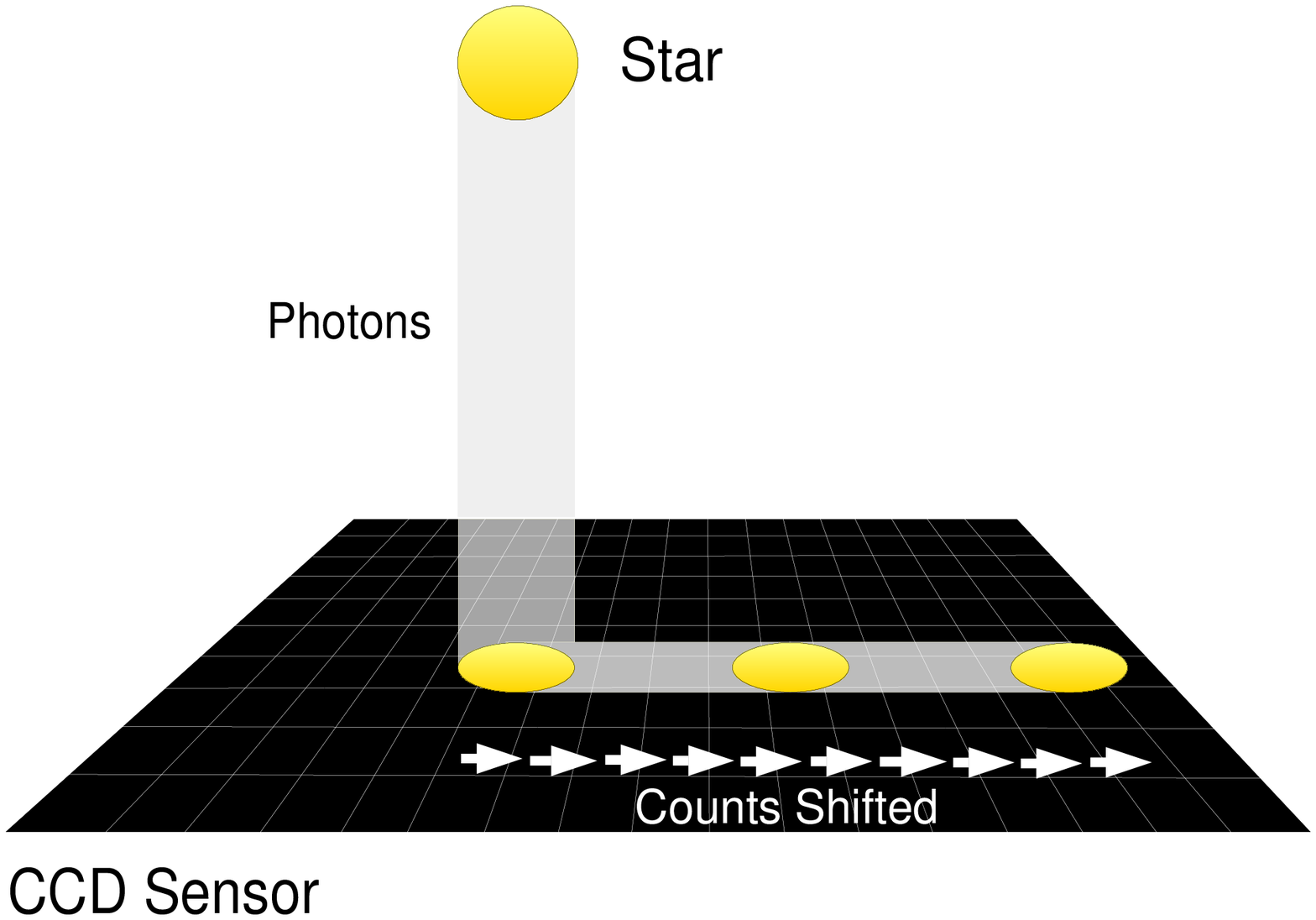,width=3.6in,angle=0}}
\caption{Zipper mode, with number of pixels between holds, $z=5$. The
figure at the top shows an earlier time than the one at the bottom.}
\label{zippic}
\end{figure}

\begin{figure}[htbp]
{\psfig{figure=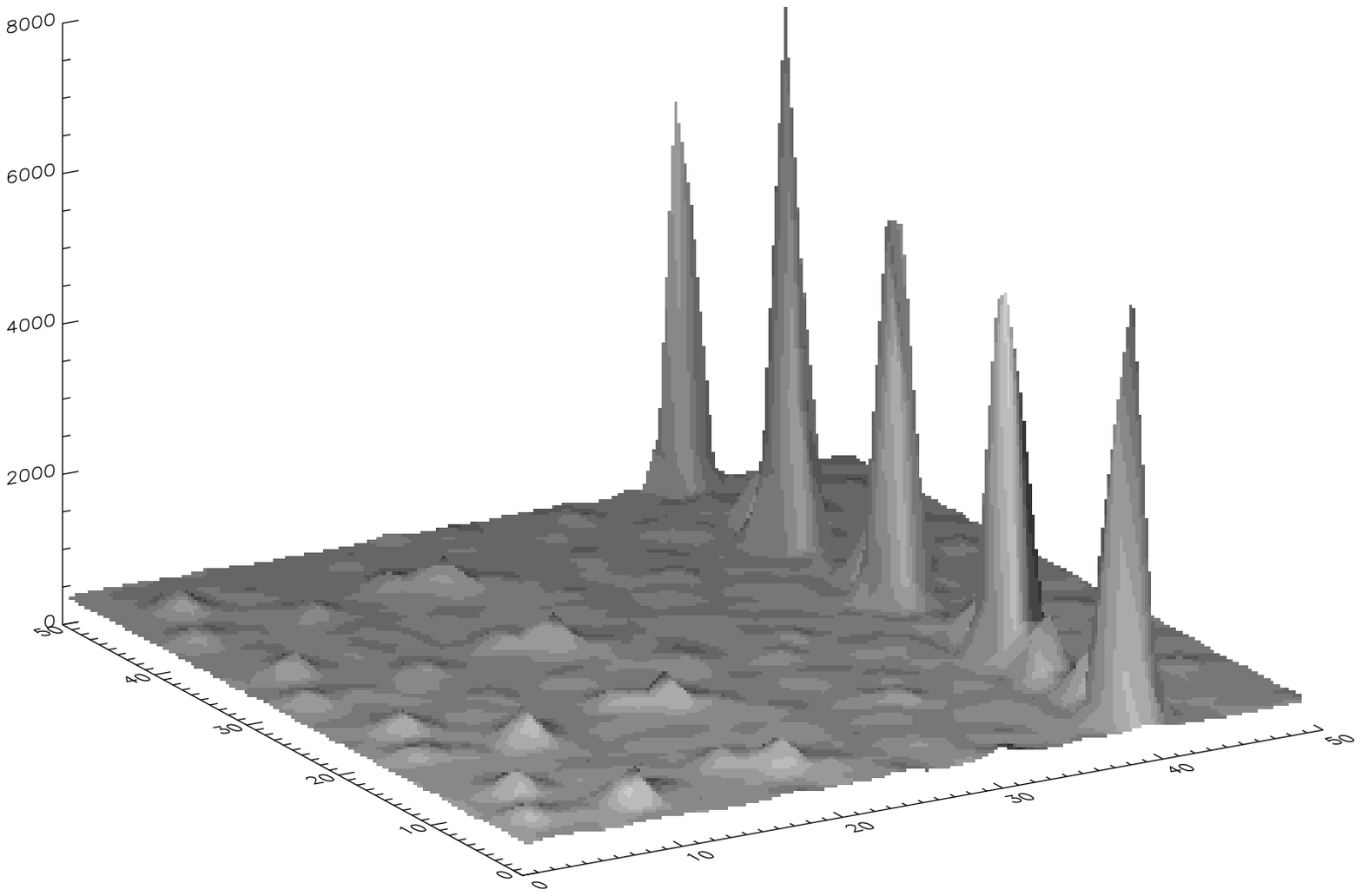,width=3.7in,angle=0}}
{\psfig{figure=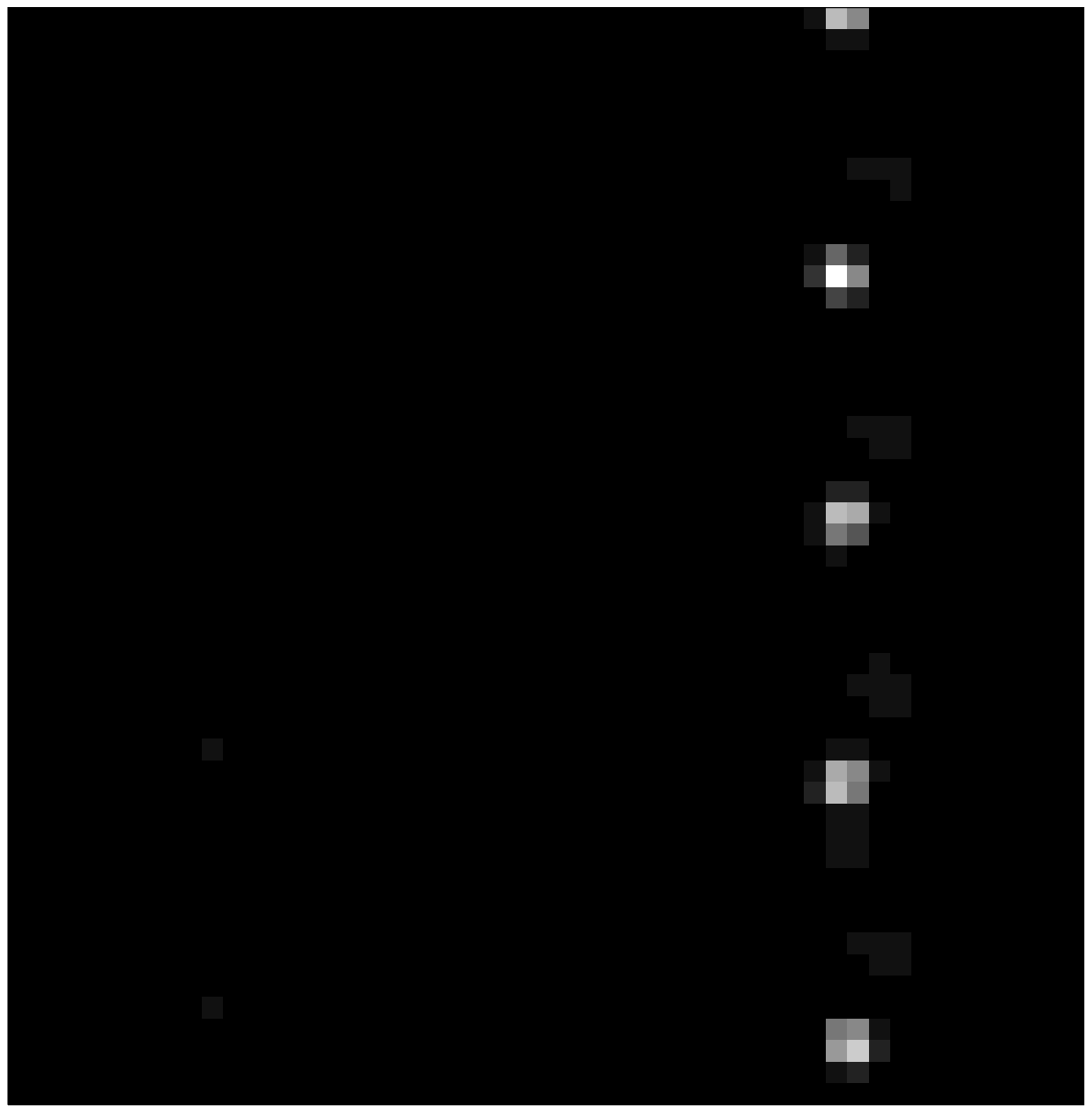,width=2.8in,angle=0}} \caption{Surface plot
and intensity plot of a section of simulated
  zipper mode image.}
\label{zipim}
\end{figure}

The proposed mode of observation for TAOS is the following: we track
the stars at the sidereal rate (the rate at which distant stars appear
to move across the sky) for $dt$ seconds, so that a star will always
illuminate the same pixels, then rapidly shift the CCD counts with
respect to the star by a set
 number of pixels, at set intervals of
time. The ``star trail''  would then resemble a
zipper, with clusters of counts at spaced intervals, which we
refer to as {\em holds}, with
 the
\emph{holdtime} $dt$. Charge from successive rows of pixels will  fall
into the horizontal register of the CCD and may be read off. The
resulting ``exposures'' are very long: data will be read out
continuously and we will only close the shutters when we wish to
monitor another star field. Figure \ref{zippic} shows the data-taking
process at two different time-points. Figure \ref{zipim} shows a
section of a simulated zipper mode image. The alternative to zipper
mode would be to hold the telescope's position fixed so that the stars
would produce trails across the CCD. A disadvantage to this latter
scheme is that the shutter would have to be frequently closed and
opened in order that the accumulated photons from the sky background
did not overwhelm the trails, with consequent frequent shutter
failures. It also turns out that detectability is higher in zipper
mode.

Smearing   resulting from the transit time between holds cannot be
ignored. The pixels  accumulate counts from the star as we wait for
each row to be read off the horizontal register of the CCD. For a
transit time of 1.3ms versus a holdtime of 200ms, the counts in the
``transit'' pixels will roughly be $1/150^{th}$ of the counts for the
``hold'' pixels. This may be a problem if we have a very bright star,
which smears over a less bright star, even if  the hold pixels do not
coincide.

The degree of overlapping stars depends on the star field being
monitored: typically the more crowded the field, the more overlapping
occurs. The amount of overlapping, for any particular mode of
observation, may be determined before observation begins. We note that,
for shutterless zipper mode, with regularly spaced holds, except for
the first few and last few observations, an overlapped hold will always
be overlapped and an nonoverlapped hold will  always be nonoverlapped.

\section{Noise and Signal} \label{noise-signal}

\subsection{Noise} \label{noise-sect}
The images contain noise from several sources. There are random
fluctuations in the photon counts, which are modeled as Poisson noise.
Atmospheric  turbulence causes photons originating from a particular
source to fall onto different locations on the CCD detector at
different points in time, giving rise to ``image motion.'' The noise
introduced by the CCD electronics (readout and dark noise) is assumed
to follow a known Gaussian distribution. Additionally there may be
stellar occultations by terrestrial (e.g. birds) and extra-terrestrial
objects (e.g. asteroids), other than KBO's. Gamma ray bursts may inject
photons; diminution and augmentation of star intensity may be due to
the star being variable and not all variable stars are cataloged.

\subsection{Signal} \label{signal-sect}
The occultation of a star by a KBO will result in a reduced photon
count from the star for the {\em duration} of the occultation, where the
{\em amplitude} of the occultation is a measure of this reduction.
The {\em onset time} is defined as the time of arrival of the
``signal,'' which is the reduction in counts from that star.

\begin{figure}[htbp]
\centerline{\psfig{figure=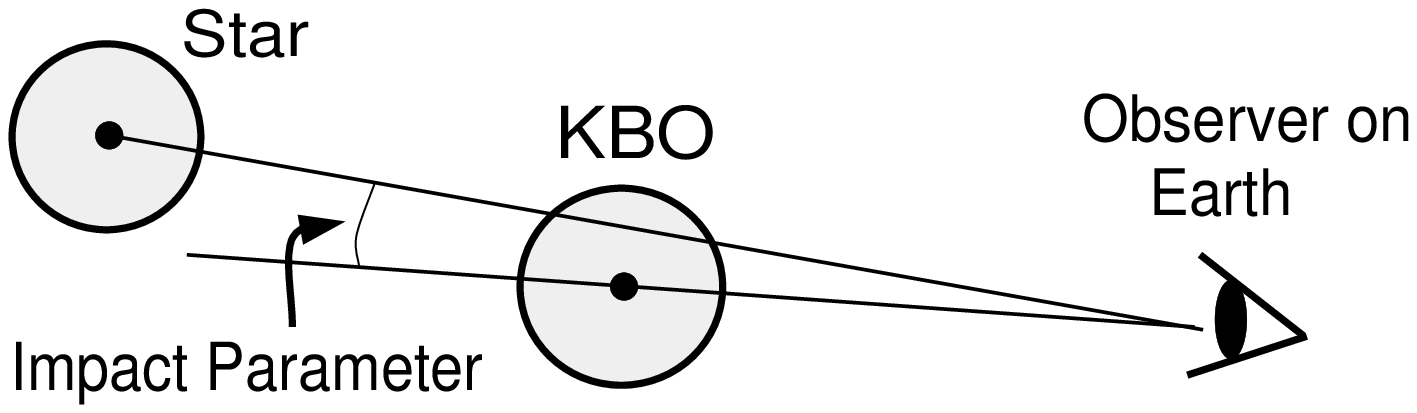,width=3.7in,angle=0}}
\caption{The impact parameter for the occultation of a star by
  a KBO.}
\label{impactpic}
\end{figure}

The signal is  affected by the diameter of the KBO and the {\em impact
parameter} of the occultation, which we define as the minimum angular
distance over the KBO's trajectory between the center of the occulting
KBO
 and the center of
the star that is being occulted. If the impact parameter is zero, the
path of the center of the KBO will cross that of the center of the
star, so that for certain angular sizes of star and KBO, the reduction
in counts from the star will be total, for at least one instance in
time---see Figure \ref{impactpic}.

\begin{figure}[htbp]
\psfig{figure=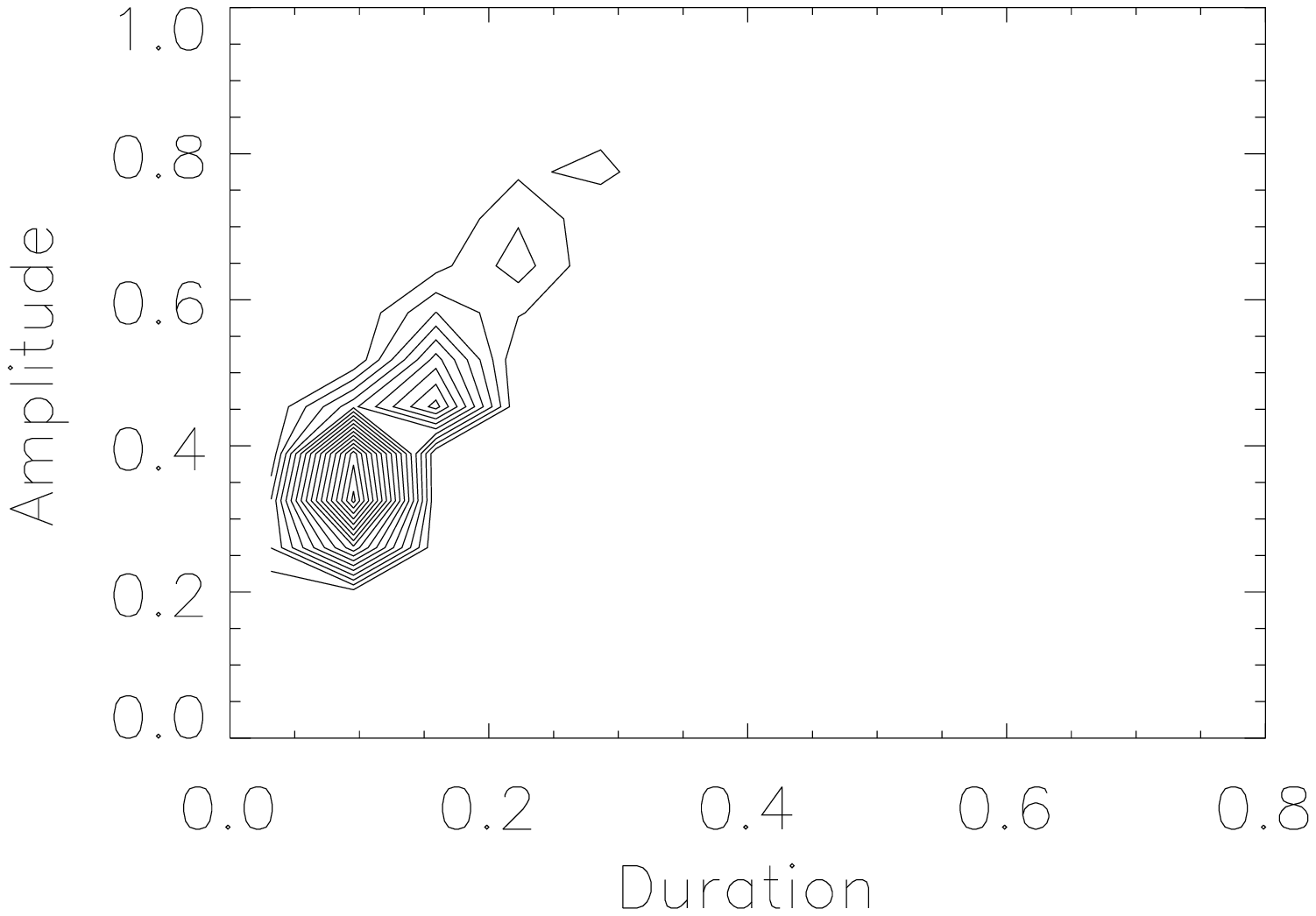,width=3.1in,angle=0}
\psfig{figure=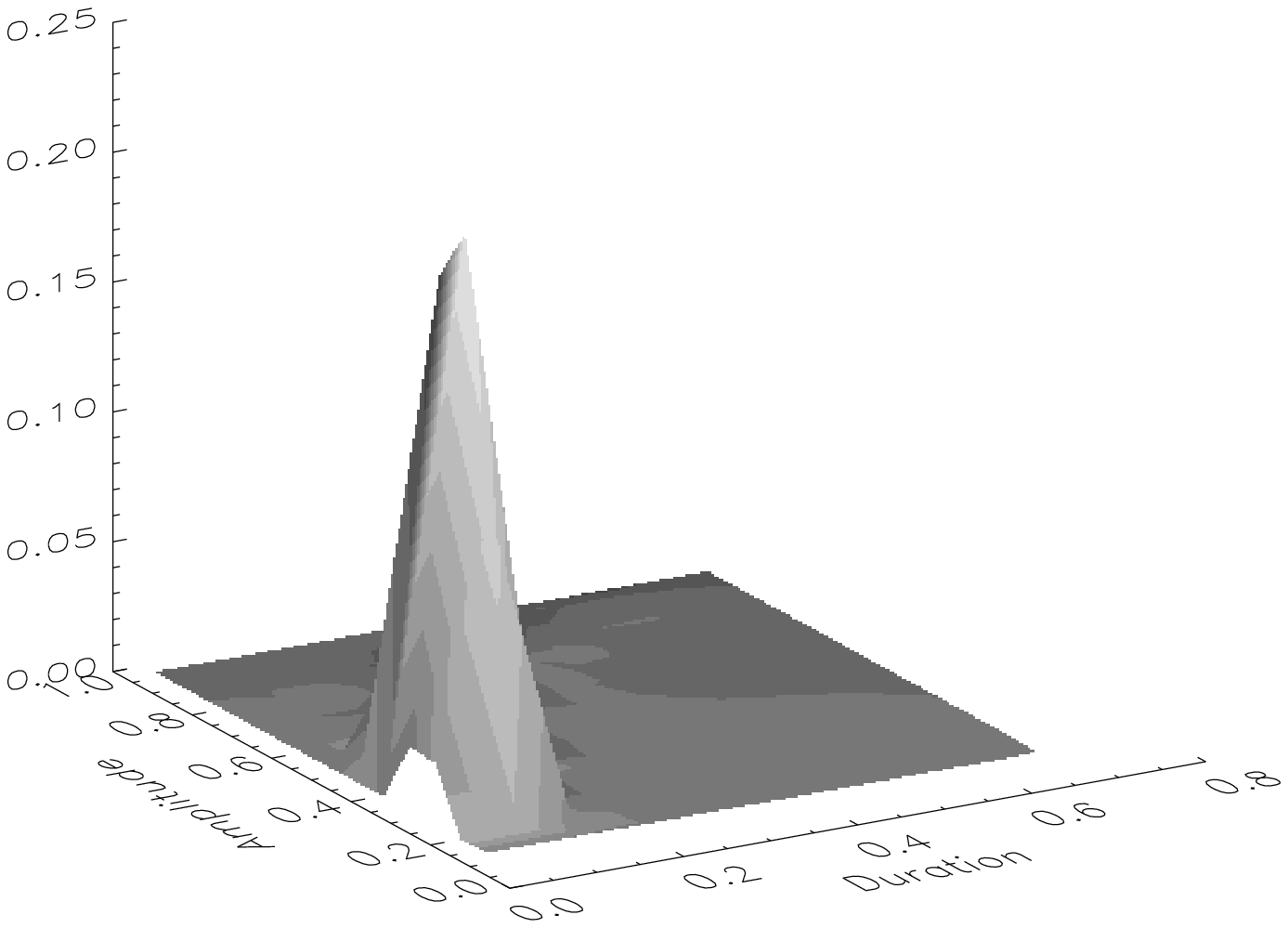,width=3.1in,angle=0} \caption{The joint
probability density function of amplitude and duration (in seconds)
when observing near opposition. The relative impact parameter is
restricted to the interval [0,.9]. } \label{joint-distn}
\end{figure}

We idealize a geometric occultation (one for which the effects of
Fresnel diffraction are negligible) as a square wave for which the time
of onset, amplitude (fractional decrease in flux),  and duration are
random.  We do not think that phenomena ignored by this idealization
significantly affect the results we report, since the test for an
occultation will be based on total flux reduction during a hold and
since shallow, brief occultations will have small detection
probability. For fixed heliocentric distance and angle of observation,
the joint probability distribution of the amplitude and duration is
determined by the impact parameter and the diameter of the KBO. We
assume that the impact parameter is uniformly distributed and that the
KBO diameter follows a power law with parameter $\omega =3.5$ equal to
that for asteroids in the asteroid belt. Let $N(c) d c$ denote the
number of KBO's with diameter between $c$ and $c + d c$. We have, for
$c$ between $c_{min}$ and $c_{max}$, the following approximation:
\begin{equation} N(c) dc  = N_0 \left(
\frac{c}{c_0} \right) ^{-\omega} dc
\end{equation}
For observations near opposition, Figure \ref{joint-distn} shows the resulting
joint probability distribution of amplitude and duration, determined
from the probability distributions of amplitude and duration.
From the figures we see that
the amplitude is typically about 0.3 and the duration is typically about
100ms.

Such probability models can be developed for a set of
fixed heliocentric distances and fixed angles of observation.
In the example above the angular size of all stars was fixed to be that of a
four kilometer object at 50AU. We assume that the velocity of an
object at each heliocentric distance is constant and equivalent to
that for an object making a circular orbit.

\section{Detection Scheme} \label{detection-sect}

We wish to design statistically and computationally efficient detection
algorithms, guided by theoretical arguments and simulations,  keeping
in mind that the algorithms should run in real time on a large number
of star trails.  The occultations are expected to be infrequent and for
credibility the false alarm rate should be controlled at a small level,
such as $10^{-12}$.  Since we expect to make $10^{10}$ to $10^{12}$
tests per year, this choice would lead to a false alarm rate of less
than one per year.

Consider the measurement of the flux from a given star in one hold. In
principle, using the idealizations above, the probability of that
measurement  can be evaluated under the assumption of no occultation
and under the assumption that an occultation has occurred. This  allows
construction of a likelihood ratio test (\cite{rice-book}) which would
be optimal if the idealizations and assumptions held. For computational
considerations we use a simple approximation to this test. For further
discussion of a likelihood ratio test and tests based on multiple holds
see \cite{mythesis}.

Consider first a single telescope.  For a set of preliminary
observations of each  monitored  star hold, we compute the median and
the interquartile range of the sequence of flux measurements. These
flux measurements could be obtained from summing up the counts in a
pixel neighborhood (aperture photometry) or by fitting the point spread
function (psf). (We are currently working on how to best extract flux
measurements for individual stars from the CCD array in real time with
the complications of overlapping and image motion.)  These initial
values are used to standardize subsequent flux measurements from that
star: the median value is subtracted from the flux and the result is
divided by the interquartile range. We use the median and interquartile
range rather than the arithmetic mean and standard deviation in order
to guard against occultations and aberrant outliers. Let $Y_{ksh}$
denote the resulting test statistic for hold $h$ of star $s$ on
telescope $k$.

We now describe a  test based on measurements at multiple telescopes
which flags a star-hold if $Y_{ksh}$ is sufficiently small at each
telescope, based on the set of preliminary observations. Suppose the
desired false alarm rate is $\alpha$ and there are $K$ telescopes.  For
each telescope we pool all the standardized $Y_{ksh}$  from the set of
preliminary observations together to form an ordered set $X_{k1} \leq
X_{k2} \leq \cdots \leq X_{kn}$ where $n$ is the total number of
star-holds. Let $m_k$ be the integer such that $m_k/(n+1)$ is closest
to $\alpha^{1/K}$. A new observation is now tested by standardizing its
measurement at each telescope as above, yielding values $Y_{1sh},
\ldots, Y_{Ksh}$ and is flagged as an occultation if $Y_{ksh} \leq
X_{km}$ at all telescopes. In words, a suitably small observation is
found at each telescope based on the initial set and new observations
are then flagged if they are smaller than each of these thresholds.

That this procedure has the desired false alarm rate $\alpha$ is due to
the following fact:  Let $Z_1, \ldots, Z_n$ be independent random
variables from the same continuous probability distribution and denote
the ordered observations as $Z_{(1)} < Z_{(2)} < \cdots < Z_{(n)}$. Let
$Z$ be another independent random variable having the same
distribution.  Then $P(Z < Z_{(m)})=m/(n+1)$. The applicability of this
result in our situation depends on some idealizations.  First, we
assume that the noise is independent between telescopes, which seems
plausible if they are sufficiently far apart. This assumption is
probably the most crucial and will have to be investigated once the
telescopes are operational. Second we assume that the noise affecting
the measurement in each star-hold is independent of that affecting
others. This is reasonable for the noise arising from photon
fluctuations and CCD electronics; it is less so for atmospheric
turbulence, but that is at least local in time and space. Third we
assume that when there is no occultation the probability distribution
of $Y_{ksh}$ does not depend on $s$ or $h$. One of the purposes of the
standardization described above is to make this latter assumption more
valid: dividing by the interquartile range attempts to compensate for
difference in variability among stars of differing magnitudes. All
$Y_{ksh}$ should then have approximately equal first and second
moments. If the flux distributions were Gaussian, for example, the
distributions of the standardized statistics would all be identical.

The time dedicated to collecting the preliminary set of observations
used for calibrating the test as above need not be great. For example,
suppose that 2000 stars are monitored with a hold time of 200ms on
three telescopes Then in 100 seconds we can collect $10^6$ star-holds
on each telescope, and to set a false alarm probability
$\alpha=10^{-12}$, the procedure requires a minimum of $10^4$
observations on each telescope


It is strictly appropriate to use this data-determined threshold only
if there is no occultation during the preliminary collection stage,
which would almost certainly be the case because of the rarity of
occultations.  However, even if there were one, the values of the
ordered observations used to form the test would only be slightly
perturbed.


There are significant advantages in this data-based method of setting a threshold.  The
alternative would be to rely on simulations, which might not be sufficiently
realistic and would furthermore have to be tailored to each star field, sky
level,  and
"seeing" (the extent to which images are blurred by light diffraction in the turbulent
atmosphere).  

As we state above, the strict validity of this method of determining a threshold
assumes that the statistics $Y_{ksh}$ have the same distribution, for fixed $k$. In
reality this will probably not be exactly the case, in particular because of the
effects of crowding.  Although our standardization takes account of differing level
and variability of the flux distributions of different stars, those distributions
might differ in their shapes in other ways as well, for example, they might
differ in skewness.  Thus it is beneficial to
 eliminate some stars
from the list of those monitored.  The criterion we have been using is
to identify in our simulations those stars that result in a large
number of false alarms. These false alarms are typically caused by a
combination of image motion and nearby bright stars. Eliminating such
stars in fact increases the chances of detecting occultations of others
since their inclusion results in a more stringent threshold.

In principle we need not combine observations from different stars but
could construct a different threshold for each star by applying the
procedure described above star by star, with no pooling.  If there were
three telescopes, we would need about 30 minutes to collect $10^4$
observations on each telescope, which would be just sufficient to
construct a test with a false alarm probability of $10^{-12}$. If there
were four telescopes only 3 minutes would be required.  An intermediate
alternative would be to group the stars into homogeneous sets.


\section{Selected Results from Simulations} \label{result-sect}

We constructed a simulator incorporating zipper mode operation and the noise
sources mentioned in Section \ref{noise-sect} and occultations generated from the
model described in Section \ref{signal-sect}. Simulated images may be used to
determine operational parameters to maximize the detection rate.

The simulator takes a star field as input; in the results described
below we used the USNO catalogue. We consider two star fields
(\ref{twofield}), which we will denote as Crowded Field (RA 4.905h, Dec
29.275degree) and Sparse Field (RA 9.61833h, Dec  0.7250degree). The
Crowded Field has 56,044 stars and the Sparse Field has 11,467 stars in
three square degrees.  The catalogue is complete down to magnitude
17-18 and the sky background brightness was set equivalent to that of a
star of of magnitude 20 (magnitude is a measure of the brightness of
stars on a logarithmic scale with larger magnitudes corresponding to
dimmer objects).

We first illustrate the advantages of using multiple telescopes
revealed by  a simulation, in which the telescopes were taken to be
identical and independent.  Table \ref{multi-telesc} shows the
detection probability for different false alarm probability and
differing numbers of telescopes, based on a simulation of 418 holds for
each of 371 stars with around 1500 occultations generated from the
distribution shown in Figure \ref{joint-distn}. From this table we see
that using two telescopes rather than one enables us to reduce the
false alarm rate from $10^{-4}$ to $10^{-8}$ with only a very slight
decrease in detection rate. Introducing a third telescope has the
effect of reducing the false alarm probability to $10^{-12}$, while
only decreasing the detection probability from 3.6\% to 3.5\%.   If the
false alarm probability is held fixed, the detection rate increases
when more telescopes are used. Of course the details of these results
depend upon the simulation and even more importantly upon operational
parameters such as choice of star field and the background level of
light in the sky. Qualitatively, the important conclusion is that by
using more telescopes, false alarm probabilities can be drastically
decreased  with little loss in detection rates and that detection
probabilities can be increased while the false alarm probability is
held fixed. Although the detection probabilities are small, we note
that there are many more dim stars than bright ones, and occultations
of the former are much harder to detect. If only bright stars were
monitored, the probabilities would be substantially larger (90\% or
more for magnitudes less than 10.5 or so), but there would be a loss in
the total number of detections. This issue is discussed further below.

\begin{table}
\caption{False Alarm and Detection Probabilities for Single and Multiple
Telescopes near Opposition. Stars with magnitude less than or equal to 15 were monitored.}
\begin{center}
\begin{tabular}{|c|c|c|c|c|c|c|c|}
\hline
\multicolumn{2}{|c|}{One Telescope}&\multicolumn{2}{|c|}{Two
Telescopes}&\multicolumn{2}{|c|}{Three Telescopes}
&\multicolumn{2}{|c|}{Four Telescopes}\\
\multicolumn{2}{|c|}{}&\multicolumn{2}{|c|}{}&\multicolumn{2}{|c|}{}
&\multicolumn{2}{|c|}{}\\
\hline
F.A.  &Detection &
F.A.  &Detection &
F.A.  &Detection &
F.A.  &Detection \\
Prob &Prob&Prob &Prob
&Prob &Prob&Prob &Prob\\
\hline
$10^{-2}$&0.187&$10^{-4}$&0.126&$10^{-6}$& 0.110 &$10^{-8}$ &0.101\\
$10^{-3}$&0.095&$10^{-6}$&0.076&$10^{-9}$& 0.060 &$10^{-12}$&0.056\\
$10^{-4}$&0.036&$10^{-8}$&0.036&$10^{-12}$&0.035 &$10^{-16}$&0.033\\
\hline
\end{tabular}\\
\end{center} \label{multi-telesc}
\end{table}

Decisions must be made about whether to observe near quadrature or
opposition. On the one hand occultations observed at quadrature will
have longer durations and thus should be easier to detect, but on the
other hand there will be more occultations at opposition because the
speed of a KBO relative to earth will be larger. Simulations can be
used to investigate this tradeoff and  Figure \ref{quad-vs-opp} shows a
comparison. The value $\mu$ on the vertical axis is proportional to the
expected number of occultations detected, assuming that the occultation
rate is proportional to the relative speed of the KBO to earth. (The
constant of proportionality depends upon the density of KBO's,  the
number of stars monitored, and other factors as well). $\mu$ depends
upon the cutoff magnitude  for monitored stars as shown on the
horizontal axis. From the figure we conclude that it is favorable to
observe near opposition and that there is little gain in monitoring
stars with magnitudes greater than 13.5. We note however, that near
quadrature it is likely  successive holds would be occulted, and a test
that took into account multiple holds would result in higher detection
rates. In one simulation this increase  was of the order of 20\%, which
leads us to conclude that overall it is still favorable to observe near
opposition.

\figurepair{\psfig{figure=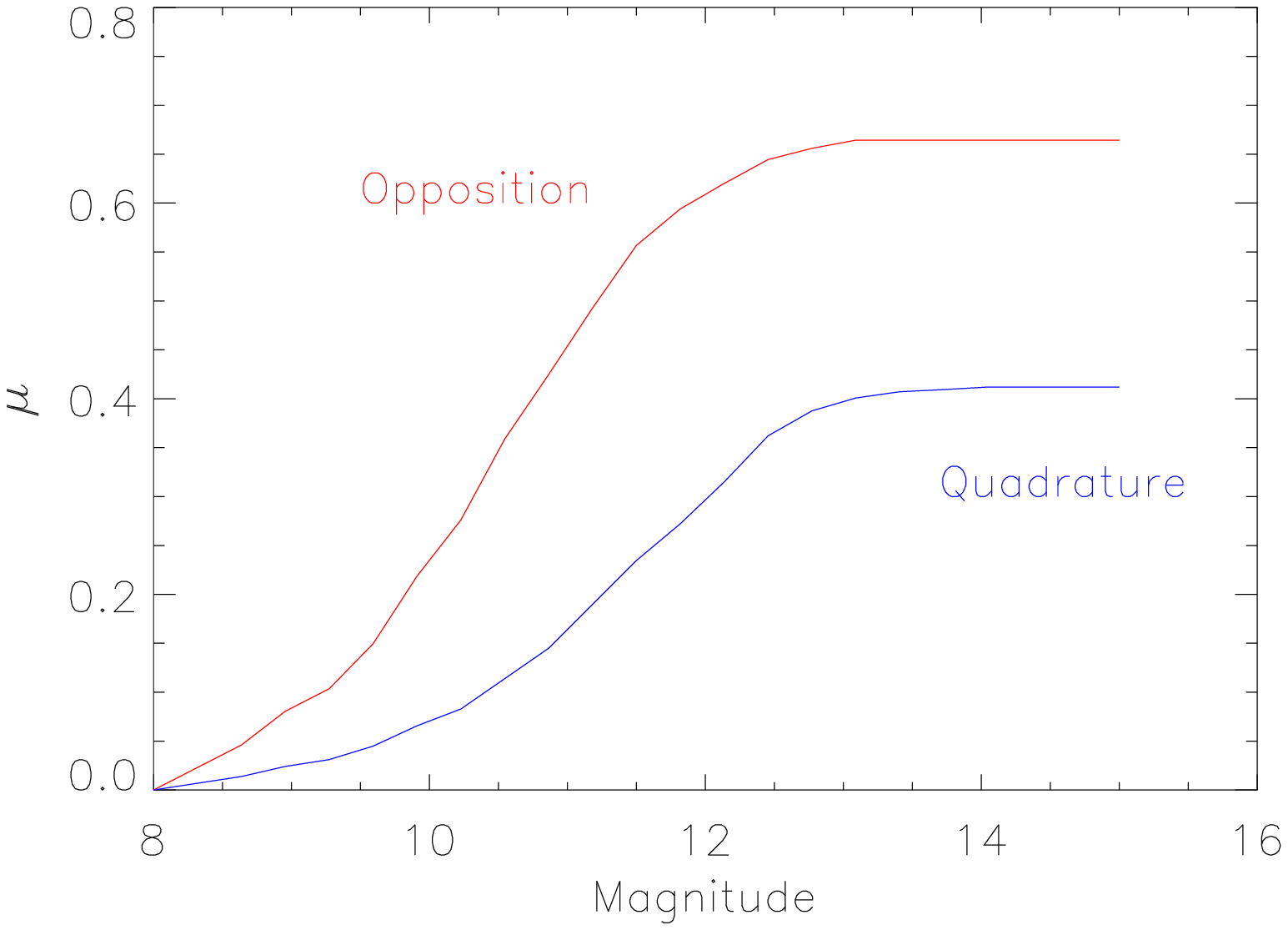,width=2.87in,angle=0}}
{\caption{Comparison of the efficacy, $\mu$, of observing near
quadrature or opposition for the Crowded Field and a hold time of
200ms.} Results are shown as a function of cut-off magnitude (i.e.,
only stars brighter than the cut-off magnitude are monitored).
\label{quad-vs-opp}} {\psfig{figure=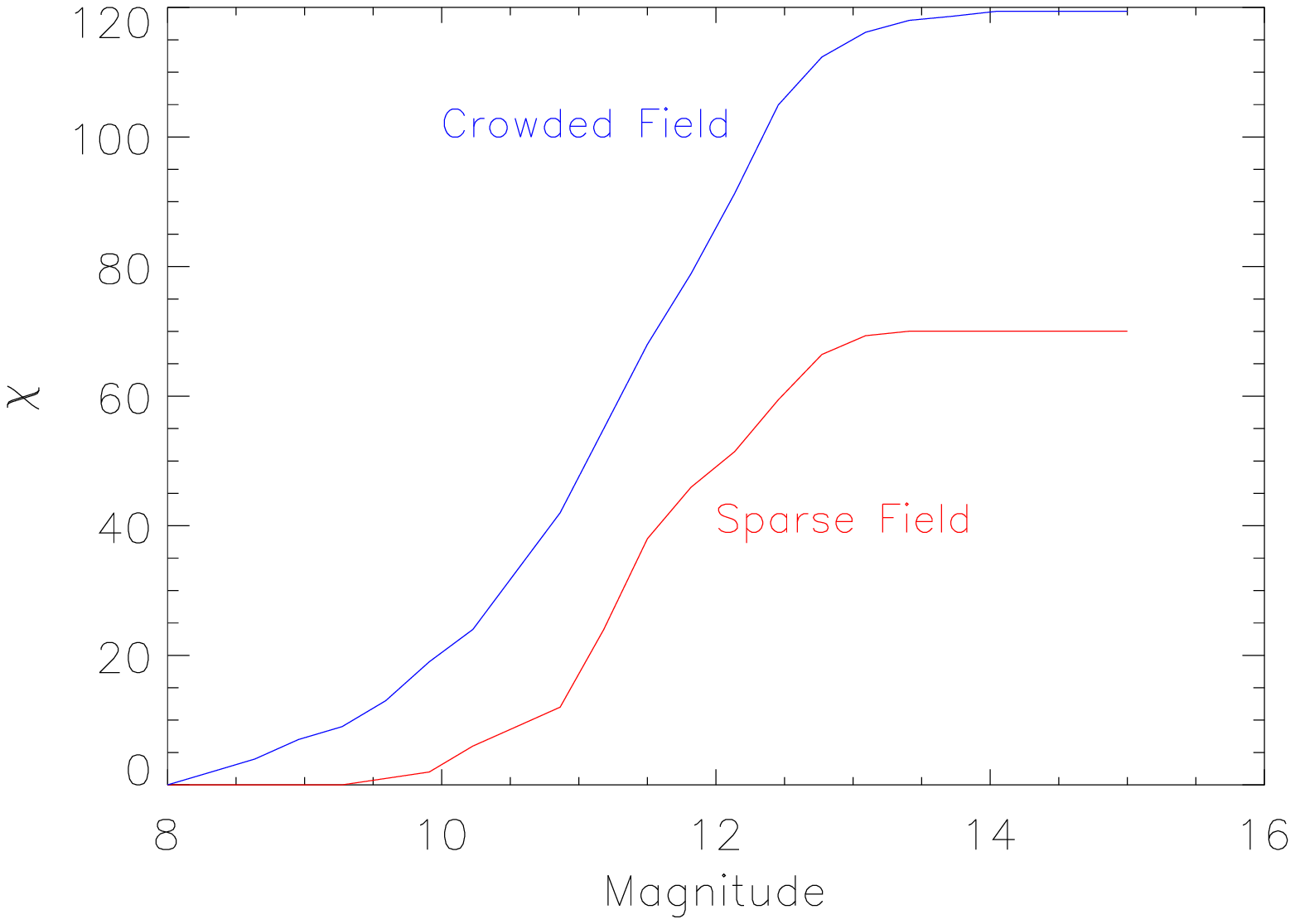,
width=3.35in,angle=0}} {\caption{Comparison of the efficacy of
observing near quadrature  for the Crowded and Sparse Fields and a hold
time of 200ms.  The value $\chi$ is proportional to the expected number
of occultations detected and is shown as a function of cutoff
magnitude.} \label{chi-plot}}

The choice of star fields to be monitored can also be guided by simulation.
Again there are tradeoffs: although crowding may increase the noise level,
especially when coupled with image motion, making
the detection of
individual occultations more difficult, a larger number of stars would be
monitored hence increasing the number of occultations per unit time. Figure
\ref{chi-plot} compares the expected number of occultations detected in a
crowded versus a sparse field. We see that there are substantial gains to be
made in monitoring the crowded field.

The simulator thus has great utility in guiding the choice of various
operational parameters. As a further example, we can examine the
effects of alternative choices of zipping parameters {the duration of
holds and the number of shifts between holds) for different angles of
observation and different phases of the moon.

\section{Retrospective Estimation of Occultation Rate}

In the previous section we focused on detecting occultations in real or
near-real time. This would enable discovery of particular objects, some
of which might be large enough to follow up optically. Beyond this
however, there is great interest in estimating the total abundance of
KBO's or at least placing bounds on this quantity, since essentially
nothing is known about the small (km sized) population. Such
information will help constrain theories of the formation of our solar
system. Thus we would wish to try to pool the results of observations
over a long period of time, such as a year, to address this question.
Note that there is a distinction between making and enumerating
detections in real time as compared to such retrospective studies.  For
example, in a retrospective mode, one is not as interested in which
particular events showed occultations as in estimating the total number
of such events. Thus, different methods of statistical analysis may be
appropriate in the two endeavors.

\begin{figure}[htbp]
\psfig{figure=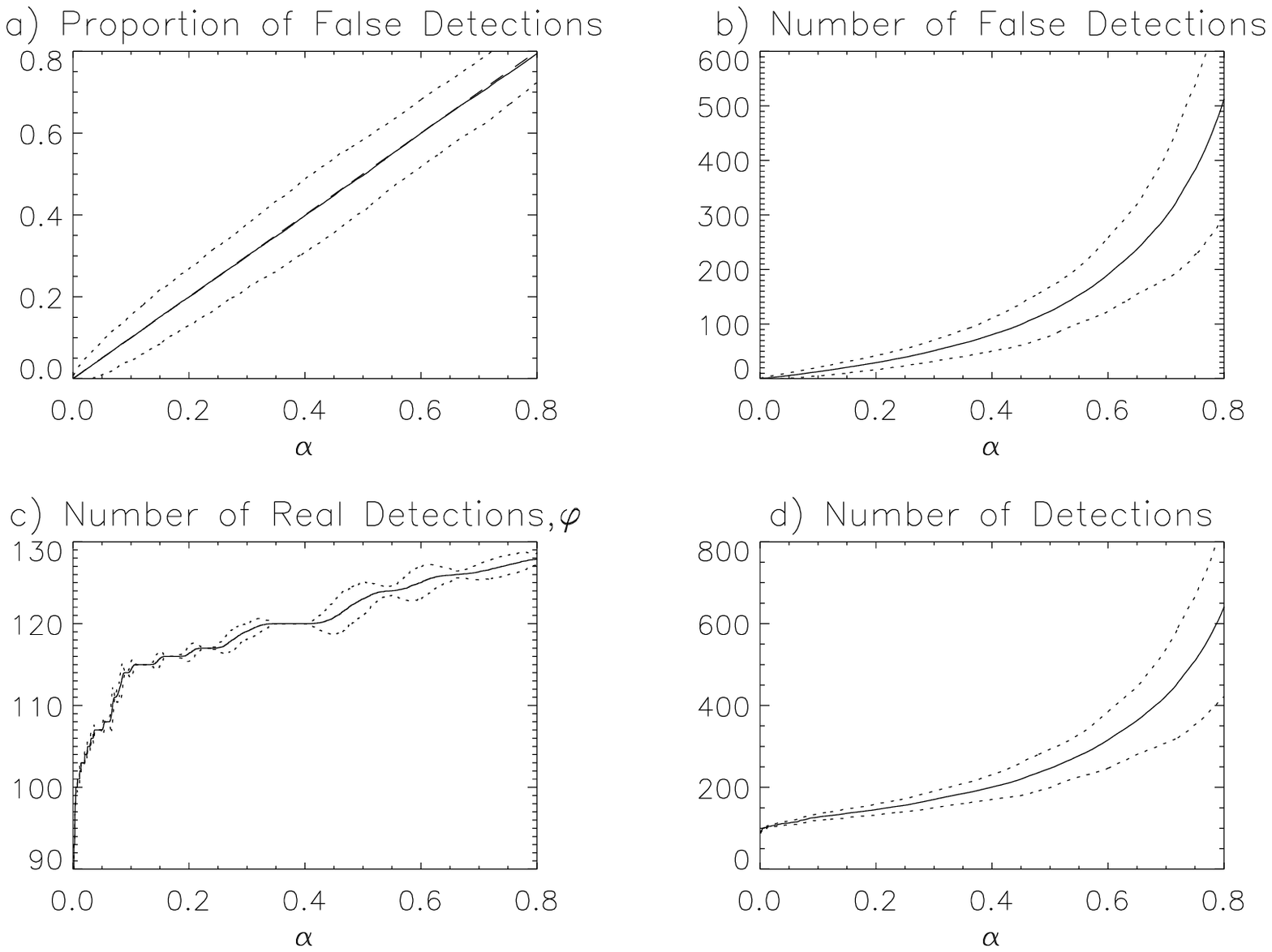,width=5.9in,angle=0} \caption{ Mean (Unbroken)
$\pm$ 2 standard deviations (Dotted), using 200 repetitions, for a)
Proportion of False
  Detections; b) Number of False Detections; c) Number of Real Detections,
$\varphi_\alpha$; d) Number of Detections. $n = 10^{10}$
 and number of occultations $=$ 300.}
\label{fourfdr}
\end{figure}

From a year of observing, we would have archived the p-values of
$10^{10}-10^{12}$ tests. In all but a tiny fraction of cases, the null
hypotheses will have been true.  This unknown fraction $\lambda$, the
occultation rate, could be as large as $10^{-7}$, but is unknown even
to an order of magnitude. The distribution of P-values under the null
(no occultation) is uniform.  The distribution of the P-values under
the alternative is not identical from test to test, but depends on the
star being monitored and other operational parameters that are not
constant. Were the marginal distribution of P-values under the
alternative known, or if it could be reliably estimated by simulation,
one could consider fitting a mixture to the empirical distribution of
all the P-values, but this is not the case. In some simulations, we
found the estimate of $\lambda$  not robust to mis-specification of the
distribution of P-values when there is an occultation.

The most promising approach to retrospective analysis that we have
found is based on the concept of the ``False Discovery Rate.'' The aim
of the testing method described earlier is to control the expected
number of false alarms per unit time by controlling the false alarm
probability to be at a desired value $\gamma$. The aim of a recently
proposed procedure \cite{ben} is different; it proposes to control the
expected proportion of flagged signals which are false alarms---the
False Discovery Rate (FDR).  The procedure works as follows.  Suppose
that $n$ independent hypotheses have been tested with corresponding
ordered p-values, $p(1) \leq p(2) \leq \cdots \leq p(n)$. In our case
these would be p-values determined as described above in section
\ref{detection-sect} for all the $n$ occultation tests conducted during
the period under consideration. Let $0 < \alpha < 1 $ be the desired
value for the FDR and let  $k$ be the largest integer such that $p(k)
\leq \alpha k/n$. Then if hypotheses corresponding to $p(1), \ldots,
p(k)$ are rejected, the FDR is less than or equal to $\alpha$. To
contrast the two approaches, note that an event is flagged by the
former procedure if it's p-value is less than $\gamma$, whereas it is
flagged by the FDR procedure depending on the rank of its p-value,
which could only be determined retrospectively.

This procedure is illustrated in the following simulation. Suppose we
monitor stars of magnitude less than or equal to 13.5 (see Figure
\ref{quad-vs-opp}) using three telescopes. For $n=10^{10}$ and 300 real
occultations, generated from the model described in Section
\ref{noise-signal}, the FDR procedure was implemented for $\alpha$
ranging from 0 to 0.8. Figure \ref{fourfdr} shows the results of
replicating this 200 times. Figure \ref{fourfdr}(a) is especially
noteworthy.  We see that FDR is very close to $\alpha$, not merely less
than or equal to $\alpha$ over the entire range of $\alpha$. Also, the
variance is small.  For comparison, if the procedure controlling the
false alarm probability to be $10^{-12}$  was used, the detection
probability in this simulation would have been 21.8\% and thus we would
expect 65 occultations to be detected. (The detection probability is
substantially higher than in Table \ref{multi-telesc} because of the
magnitude cut-off.) By contrast, if FDR were used to determine a subset
with $\alpha$ equal to .05 or .10, we see from Figure \ref{fourfdr}(c),
that more than 100 real occultations would be included.

Figure \ref{fourfdr}(a) suggests that if FDR is controlled at level
$\alpha$, the proportion of false detections will in fact be quite
close to $\alpha$. Thus if the number of flagged signals is
$\Omega_\alpha$, it is natural to estimate the number of real
detections among them by $\hat{\varphi}_\alpha = (1-\alpha) \times
\Omega_\alpha$. Figure \ref{phi-plot} shows the error in estimating
$\varphi_\alpha$ as a function of $\alpha$. We see there that the error
is small, but increases as $\alpha$ increases.

As noted above, the fraction of star-holds in which occultations
occurred would provide important information about KBO abundance. We
thus propose a method for estimating the occultation rate.

\figurepair{\psfig{figure=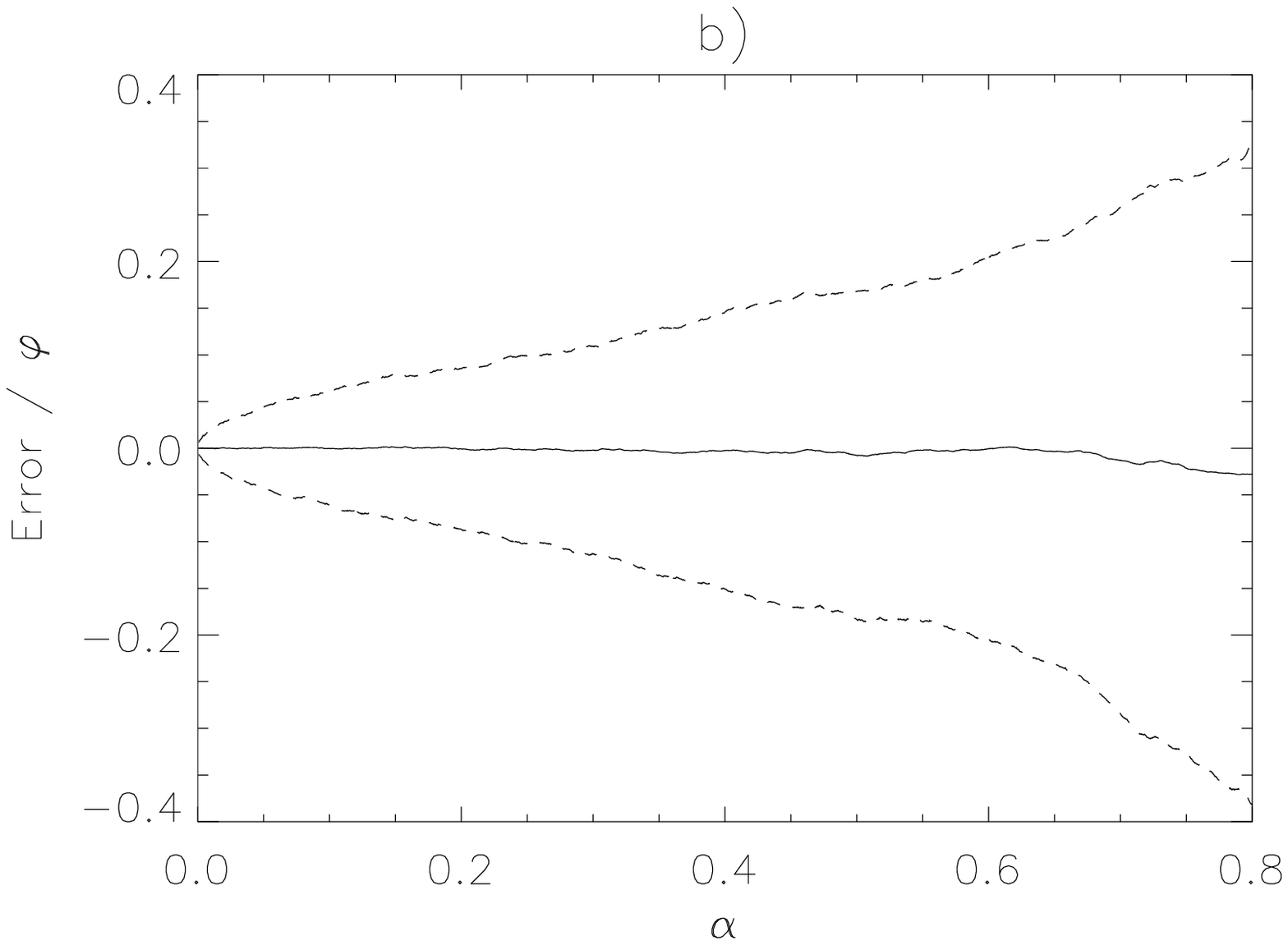,width=3.1in,angle=0}}
{\caption{The Proportional Error  (mean plus or minus 2 SD's) in
Estimates of $\varphi_\alpha$ in 200 Repetitions. $n = 10^{10}$
 and number of occultations $=$ 300.}
\label{phi-plot}} {\psfig{figure=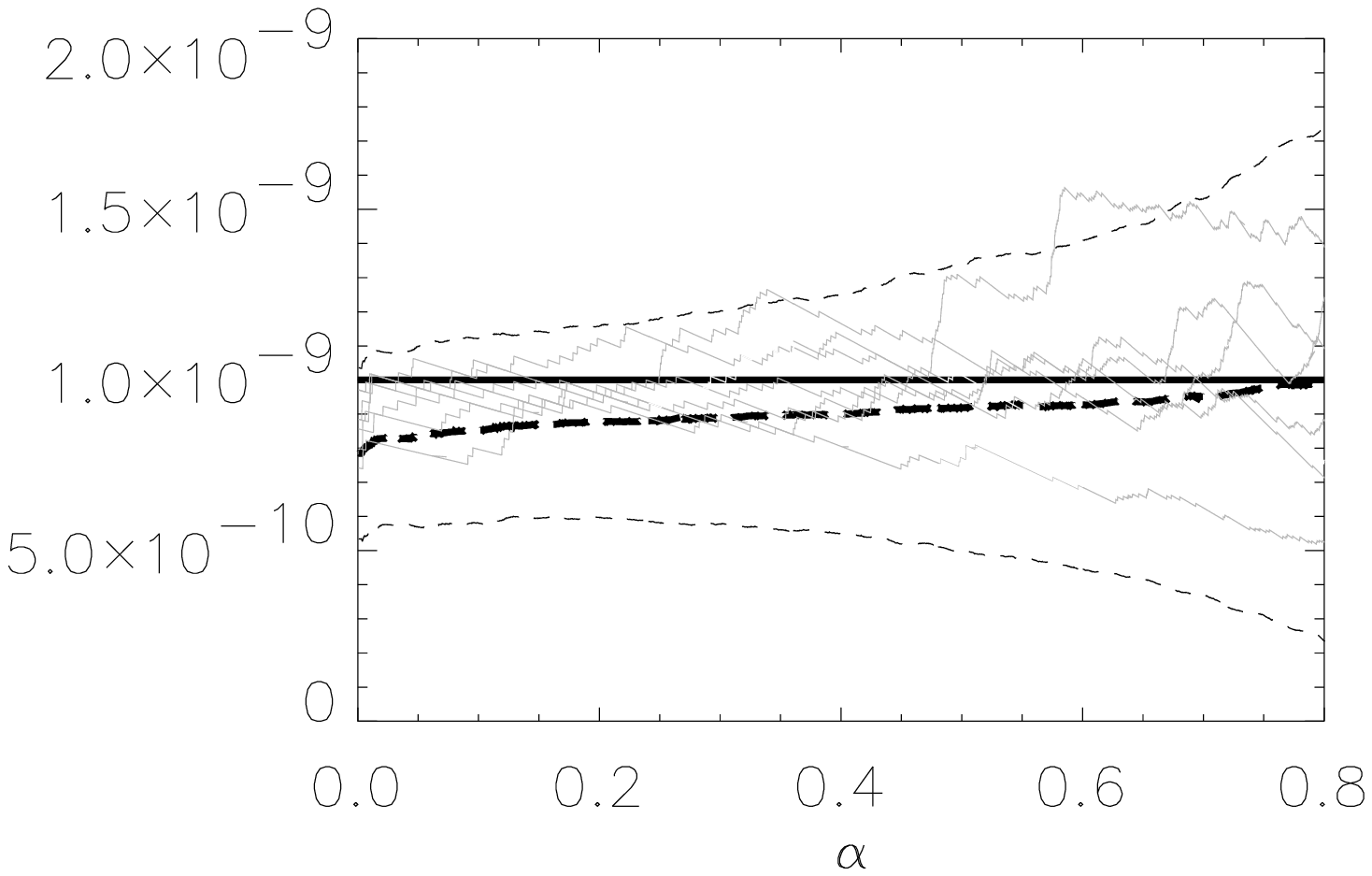,width=3.1in,angle=0}}
{\caption{The average value (bold dash) plus and minus two SD's
   (faint dash)
  for $ \hat{\lambda}_\alpha$ versus $\alpha$, when
  $\lambda = 10^{-9}$(bold unbroken) , together with
$\hat{\lambda}_\alpha$ in 6 simulations (light grey lines). Number of Tests = $10^{11}$.
A cutoff magnitude of 11 was used.}
\label{ed4lam}}

We note that the number of occultations will be greater than or equal to the
number of real detections.  An estimated lower bound for the number of occultations will
then be given by $\hat{\varphi}_\alpha$. We define $\lambda$ to be the
occultation
rate in our data, given by the ratio of the number of occultations to the total
number of tests conducted. In this way, lower bounds for $\lambda$ as a function
of $\alpha$ are estimated by $\hat{\lambda}_\alpha =
\hat{\varphi}_\alpha/n$. The number of real detections will
increase with $\alpha$ so, viewing $\hat{\lambda}_\alpha$ as an estimate
of $\lambda$, the bias will decrease as $\alpha$ increases.  On the other hand,
the variance increases with increasing $\alpha$ as shown in Figure
\ref{phi-plot}.  Both these effects are shown in Figure \ref{ed4lam}.

As noted previously, the brighter the star, the higher the detection
probability in general, so it may be beneficial to limit estimates to
brighter stars for the purpose of estimating the occultation rate.  In
Figure \ref{fdrdifflam11b123} we see that as $\lambda$ increases, the
relative error decreases. When only relatively bright stars are used,
it appears that smaller values of $\alpha$ result in superior
performance. However, if fainter stars are used as well, the optimal
value of $\alpha$ seems to depend on the number of tests and the value
of $\lambda$.  Overall, it is encouraging that the RMS error is not
terribly sensitive to the value of $\alpha.$

\begin{figure}[htbp]
\psfig{figure=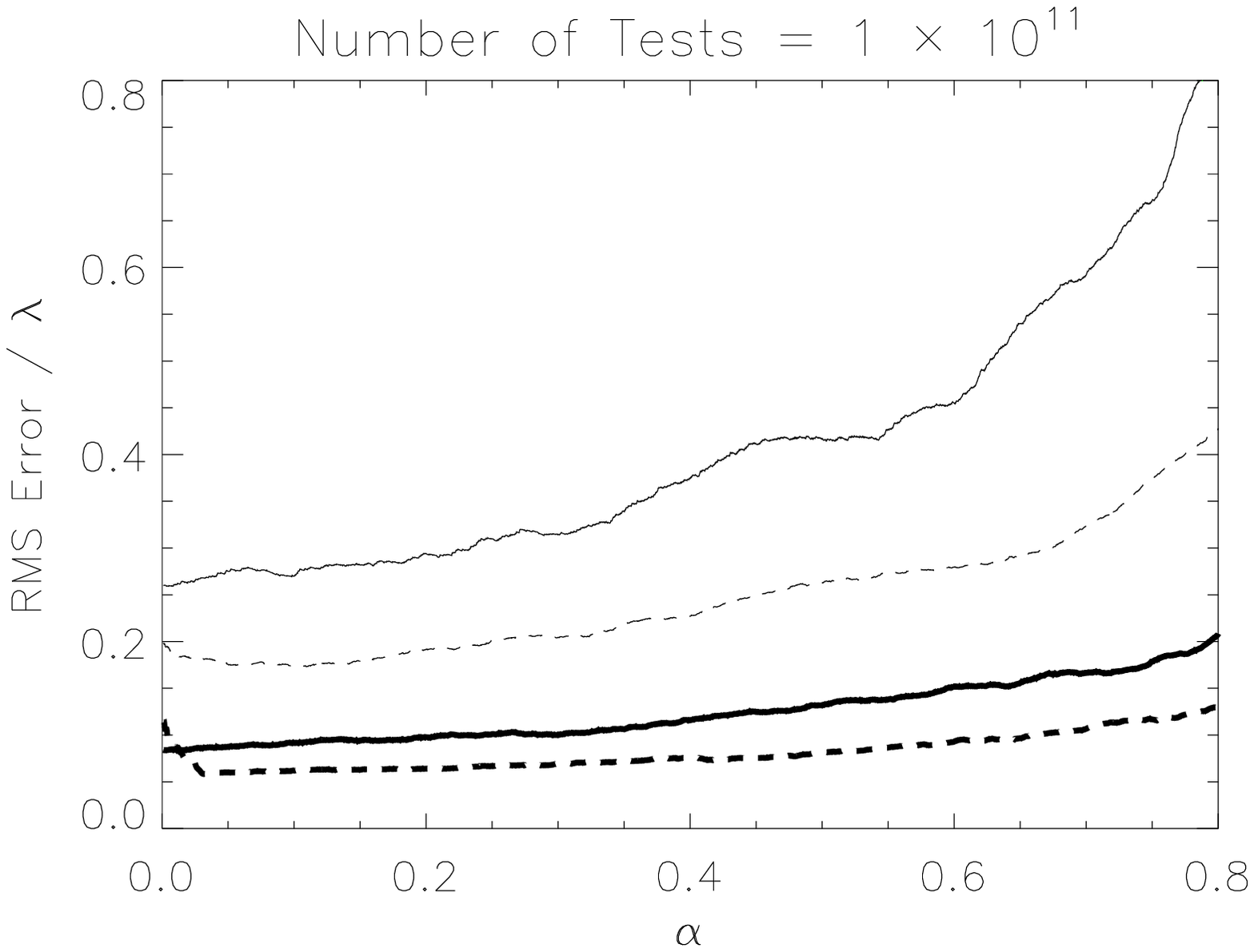,width=3.1in,angle=0}
\psfig{figure=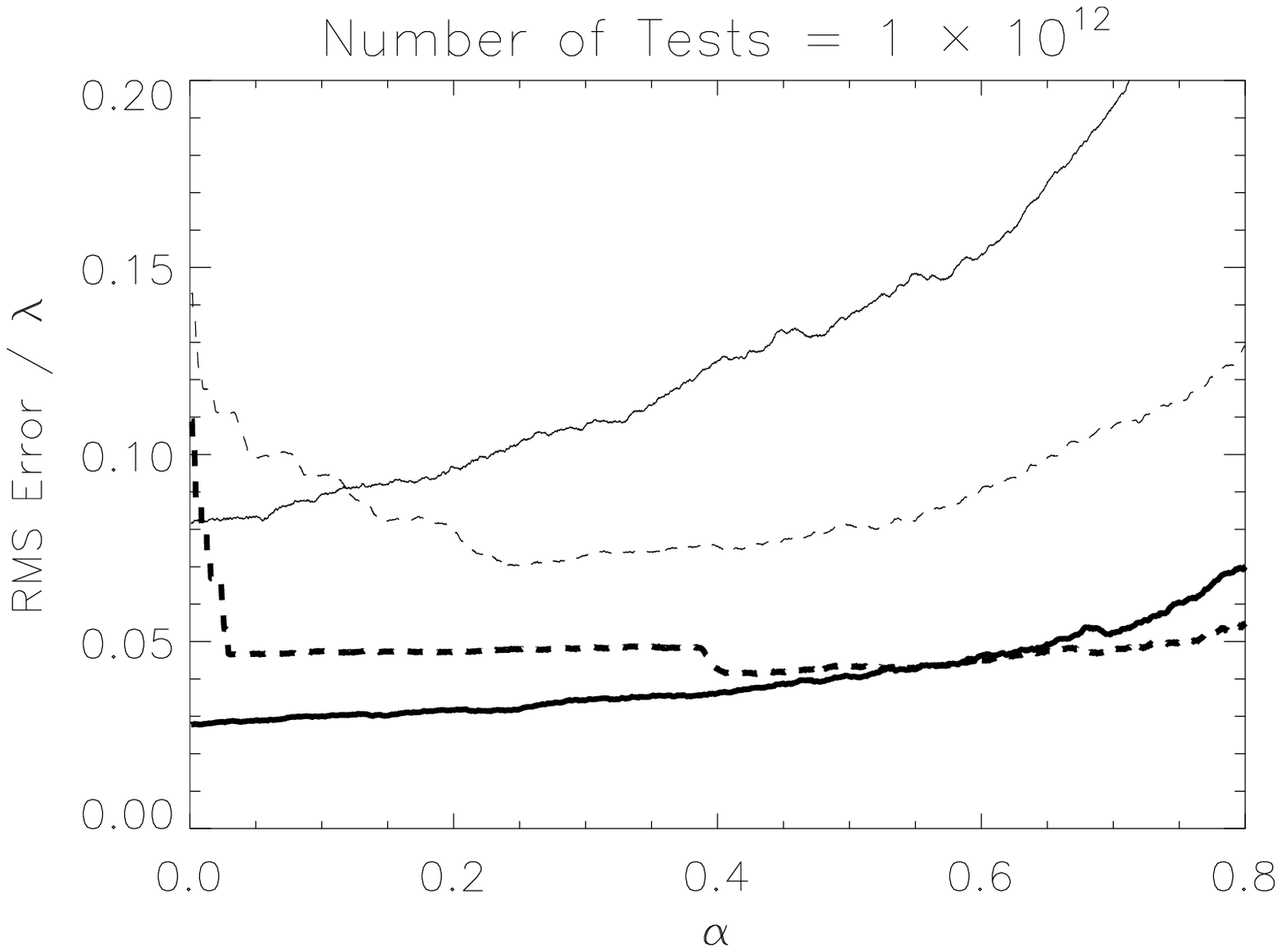,width=3.1in,angle=0}
\caption{Relative RMS error in estimating $\lambda$. Number of Tests,
near quadrature: (left) $1 \times 10^{11}$, (right) $1 \times
  10^{12}$. Bold: $\lambda= 5 \times 10^{-9}$, faint:  $\lambda=5 \times 10^{-10}$.
  Cut-off magnitude:
  (unbroken) 11.0,
   (dashed) 12.1. }
\label{fdrdifflam11b123}
\end{figure}

\section{Conclusion}
\label{Conclusion}

We have shown how multiple telescopes may be used to control the false
alarm rate for real-time detection, by using thresholds determined in a
data-based manner. Realistic simulations illustrated the effectiveness of
this procedure and may be used to guide the choice of various operational
parameters.  We have argued that once a substantial body of data has been
gathered, an alternative method based on the FDR procedure can be used to
estimate the occultation rate.  Simulations indicate that this method can
produce useful results. It is important to note that different statistical
procedures are appropriate for real time detection as compared to retrospective
analysis.

The data archive resulting from TAOS will be a valuable resource for
other astronomical research as well.  For example, little is known
about stellar variability on a sub-second level and TAOS will provide a
vast collection of such time series. There may well be unexpected
phenomena in the archive and in order not to miss them we face a
problem of searching for needles in a haystack when we don't know what
needles look like. How can we automate serendipity?

\newpage

\bibliography{biblio2.bib}

\end{document}